\documentclass[A4paper,conference]{IEEEtran}
\IEEEoverridecommandlockouts
\usepackage{cite}
\usepackage{amsmath,amssymb,amsfonts}
\usepackage{graphicx}
\usepackage{xcolor}
\usepackage{bm}
\usepackage[noend]{algpseudocode}
\usepackage{lipsum}
\usepackage{algorithm}
\usepackage[caption=true,font=scriptsize,labelfont=rm,textfont=rm]{subfig}
\usepackage{multirow}
\usepackage{setspace}

\DeclareMathOperator*{\argmax}{argmax} 



\def\BibTeX{{\rm B\kern-.05em{\sc i\kern-.025em b}\kern-.08em
    T\kern-.1667em\lower.7ex\hbox{E}\kern-.125emX}}

\graphicspath{{./figure/}}

\begin{document}

\title{Low Overhead Beam Alignment for \\ Mobile Millimeter Channel \\ Based on Continuous-Time Prediction}

\author{\IEEEauthorblockN{Huang-Chou Lin}
\IEEEauthorblockA{\textit{Department of Electrical Engineering} \\
\textit{National Cheng Kung University}\\
Tainan, Taiwan \\
\texttt{Q36091017@gs.ncku.edu.tw}}
\and
\IEEEauthorblockN{Kuang-Hao (Stanley) Liu}
\IEEEauthorblockA{\textit{Institute of Communications Engineering} \\
\textit{National Tsing Hua University}\\
Hsinchu, Taiwan \\
\texttt{khliu@ee.nthu.edu.tw}}
}

\maketitle
\begin{abstract}
In millimeter-wave (mmWave) communications, directional transmission based on beamforming is important to compensate for high pathloss. To maintain the desired directional transmission gain, one common method is beam scanning, which involves the transmitter sending the pilot signal over all available beam directions to find the optimal one. Alternatively, beam tracking using statistical models and kalman filter (KF) can save beam training overhead. Unfortunately, existing beam tracking solutions are limited to a fixed beam variation pattern. In this work, we propose an adaptive online beam alignment (AOBA), which aims to reduce beam alignment overhead and achieve accurate beam alignment for any movement profile. The proposed AOBA periodically performs beam tracking using a small amount but carefully selected candidate beams and switches to beam scanning using all available beams based on a given switching rule. During the interval without the pilot signal, the optimal beam at an arbitrary time instant is predicted with the aid of the recently proposed ordinary differential equation (ODE)-long short-term memory (LSTM) model. Extensive simulations are conducted to evaluate the performance of the proposed AOBA in comparison with several existing beam alignment schemes.
\end{abstract}


\section{Introduction}

Millimeter Wave (mmWave) communication is a promising technique to fulfill the ever-growing demand for high data rate transmission~\cite{Rappaport2013}. However, mmWave channels suffer high propagation loss and blockage effects, making it challenging to maintain seamless mmWave connectivity~\cite{Niu2015}. Fortunately, the small wavelength of mmWave allows densely packed antenna elements in a small form factor. Hence, the array antenna consisting of many elements has been commonly used in mmWave communication to achieve directional beamforming~\cite{Roh2014}. 

The alignment of highly directional and narrow beams is essential to maintain the desired beamforming gain. A standard solution to find the best beam direction is to transmit the pilot signal in all the possible directions, known as beam scanning or sweeping\cite{3GPP2018}. When the optimal beam direction changes frequently, e.g., in mobile applications, beam scanning causes dramatic training overhead. To maintain accurate directivity with low training overhead for mmWave communication, beam tracking~\cite{Bae2017} and beam prediction~\cite{Lim2021,Kaya2021,Urakami2022,Ma2023,Liu2023} have been extensively studied. Beam tracking attempts to find the optimal beam by tracking the variations of the time-varying channel gains and angles. A widely used beam tracking method is Kalman filter (KF), which estimates the channel state for each slot between two consecutive pilot transmissions~\cite{Va2016,Liu2019,Larew2019}. Thanks to periodic pilot transmissions, the training overhead of beam tracking can be significantly reduced compared to beam scanning. Further overhead reduction is possible if some candidate beams are selected to send the pilot signal~\cite{Ma2021a,Echigo2021}. 

On the other hand, beam prediction finds the optimal beam based on periodically received pilot signals. Unlike beam tracking, beam prediction does not explicitly estimate channel gain and angle variations but forecasts the optimal beam on a per-slot basis. Some existing works treat beam prediction as a time-series forecasting problem, which is a well-known problem with a variety of solutions, e.g., statistical modeling and machine learning (ML) based methods (see~\cite{QurratulainKhan2023} and the references therein).

While prior works have shown the success of ML-based methods in carrying out beam prediction~\cite{Kaya2021,Urakami2022,Ma2023,Liu2023}, they all make the prediction based on the pilot signal transmitted over all beams, resulting in high overhead. The importance of our work lies in reducing the overhead of beam prediction for mmWave. Specifically, we propose to make predictions using partial beams, which we call the probing beams. The selection of partial beams heavily affects the prediction accuracy, and the prediction accuracy degrades with time. To overcome the former, we consider three training beam selection strategies. For the latter, we propose switching from beam prediction to beam scanning either periodically or adaptively. In this way, we can maintain accurate beam alignment by alternating different operation modes, namely, beam scanning, beam tracking, and beam prediction, while keeping the training overhead as low as possible. 

The remainder of this paper is organized as follows. Sec.~\ref{sec: model} explains the system model and problem under study. The proposed approaches to reduce the beam training overhead for mmWave are presented in Sec.~\ref{sec: proposed}, including the training beam selection schemes, the beam prediction method, and the switching rule between different beam alignment modes. Simulation results are shown in Sec.~\ref{sec: results} followed by concluding remarks given in Sec.~\ref{sec: conclusion}.

\section{System Model and Problem Description}\label{sec: model} 

Consider a mmWave network where one base station (BS) serves a user equipment (UE). The BS has $M$ antennas in a uniform linear array (ULA), and the UE has a single antenna. We focus on the analog beamforming where each BS antenna element connects to a phase shifter and all the phase shifters are connected to a single RF chain. In practice, the phase shifters have finite resolutions. This work assumes each phase shifter can vary its phase in $Q$ quantization levels.

Since the mmWave channel is sparse by nature, we adopt a geometric channel model to characterize the channel between the BS and the UE at time $t$ as~\cite{Lim2021,Ma2023}
\begin{equation}
\mathbf{H}_t = \sum_{l=1}^L \alpha_{t,l} \mathbf{a}(\phi_{t,l})
\end{equation}
where $L$ is the number of paths, $\alpha_{t,l}$ and $\phi_{t,l}$ are the path gain and angle of departure (AoD), respectively, of the $l$th path at time $t$, and $\mathbf{a}(\phi_{t,l})\in \mathbb{C}^M$ denotes the steering vector of the ULA at the BS given by $\mathbf{a}(\phi_{t,l})=1/\sqrt{M}[1~e^{j2\pi d\sin(\phi_{t,l})/\lambda}~\cdots~e^{j (M-1)2\pi d\sin(\phi_{t,l})/\lambda}]^T$ with $d$ and $\lambda$ being the antenna spacing and the wavelength, respectively. Let $s \in \mathbb{C}$ denote the transmit signal from the BS, which uses the analog beamforming vector denoted as $\mathbf{f} \in \mathbb{C}^{M}$. At time $t$, the received signal at the UE is given by
\begin{equation}
\mathbf{y}_t^{(q)} = \mathbf{H}_t^T \mathbf{f}_q s + \mathbf{z}_t
\end{equation}
where $|s|=1$ and $\mathbf{z}_t$ is the additive white Gaussian noise (AWGN) with zero mean and variance $\sigma^2$. Here, $\mathbf{f}_q$ is the analog beamforming vector selected from a codebook based on the discrete Fourier transform (DFT). The codebook contains $Q$ codewords with the $q$-th codeword given by~\cite{Ma2023,Liu2023}
\begin{equation}
\mathbf{f}_q = \sqrt{ \frac{1}{M} } [1~e^{j 2\pi \theta_q}~\cdots~e^{j 2\pi (M-1) \theta_q}]
\end{equation}
where $\theta_q=q/Q$.

Due to UE movement, the AoD of the BS-UE channel changes with time, and thus, the BS needs to update its beamforming vector to align with the AoD. To align the beam, the BS selects the optimal beamforming vector from the codebook that maximizes the beamforming gain defined as $G_t = | \mathbf{H}_t^T \mathbf{f}_q |^2$. Thus, the best beam index $q_t^*$ at time $t$ is given by
\begin{equation}\label{eq: best-beam-index-scanning}
q^*(t) = \argmax_{q\in \{1,2, \cdots, Q\}} G_t.
\end{equation}
While $q^*(t)$ can be found by \textit{beam scanning}, i.e., the BS sends the pilot signal using all available beamforming vectors and the UE measures their corresponding beamforming gains, this approach induces significant training overhead and delay when $Q$ is large. On the other hand, the DFT codebook of size $Q$ has an angular resolution of $1/Q$. As a result, $Q$ should be sufficient to ensure the angular resolution of the DFT codebook. One way to strike a balance between the angular resolution and beam training overhead is to keep $Q$ large but perform beam scanning periodically instead of every time slot. The optimal beam scanning period depends on the UE moving speed, which is often time-varying, and thus, it is not easy to optimize the beam scanning period in real time. Our remedy is to predict $q_t^*$ for the time in the fixed beam scanning period. It is worth mentioning that the BS can choose to probe a subset of available beamforming vectors instead of all available beams to reduce the training overhead further. In this case, the beam prediction accuracy relates to which beams to probe. Denote $S$ the number of probing beams to send the pilot signal. We refer to the case of $S<Q$ as \textit{beam tracking} to be not confused with beam scanning using $Q$ probing beams. 


\section{Overhead Reduction with Adaptive Beam Alignment }\label{sec: proposed}

\begin{figure}[!t]
\centering
{
\includegraphics[width=0.9\linewidth]{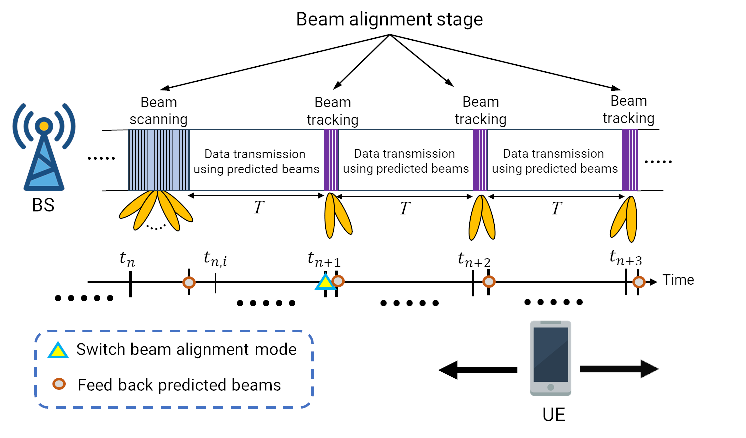}
}
\caption{Time-domain operation of the proposed beam alignment scheme.}
\label{fig: timeline}
\end{figure}

To reduce overhead for mmWave beam alignment, we consider a periodic beam alignment protocol as shown in Fig.~\ref{fig: timeline}. The protocol alternates between two stages, namely the beam alignment stage and the beam prediction stage. In the beam alignment stage, the BS transmits the pilot signal to the UE for beam alignment. The beam alignment stage operates periodically with a fixed period $T$, during which the BS transmits data to UE. On the other hand, the duration of the beam alignment stage varies depending on the operation mode, i.e., beam scanning using all available beams to transmit the pilot signals or beam tracking that considers partially selected beams as the candidate beams to probe. The candidate beam selection rule will be specified in Sec.~\ref{sec: candidate-beam}, and the mode switching criterion will be presented in Sec.~\ref{sec: mode-switching}. Based on the received pilot signal in the beam alignment stage, the UE predicts the optimal beams in the beam prediction period of length equal to $T$. The beam prediction presented in Sec.~\ref{sec: prediction} aims to accurately predict beam variations without invoking the pilot transmission, thereby reducing overhead for beam alignment. Then the UE reports the predicted beam indices to the BS, which uses the predicted beams for subsequent data transmission. In practice, the user feedback can be done through the uplink control channel using the same beam reporting procedure defined in 3GPP NR standard~\cite{3GPP2018}. Overall, the proposed scheme, the adaptive online beam alignment (AOBA), consists of three working modules, including the candidate beam selection module, the beam prediction module, and the mode switching module, as explained in the sequel.

\subsection{Candidate Beam Selection}\label{sec: candidate-beam}

From the discussion in Sec.~\ref{sec: model}, the beam alignment overhead is dominated by the frequency of beam alignment and the amount of the pilot signal. Simply reducing the beam alignment frequency or the amount of pilot signal could damage beam alignment accuracy. Since the beam variation is correlated in time and space, the current optimal beam will likely be close to the past one. Bearing this in mind, we attempt to find the optimal beam by considering the position of the optimal beam in the past. As will be shown later, this is helpful to save the beam training overhead and simplify the beam prediction complexity.

\begin{figure}[!t]
\centering
{
\includegraphics[width=0.7\linewidth]{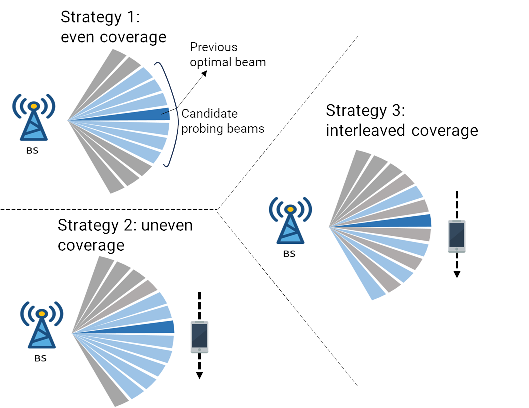}
}
\caption{Illustration of three beam selection strategies.}
\label{fig: beamselection}
\end{figure}

Our strategy is to consider the optimal beam found in the previous beam alignment stage, denoted as $q_{n-1}^*$, when selecting some candidate beams for the $n$-th beam alignment stage. We propose three strategies to determine the candidate beam set denoted by $\mathcal{S}$. Fig.~\ref{fig: beamselection} illustrates these three strategies.
\begin{itemize}
\item Even coverage: The candidate beams are selected to cover the angular domain centered at the angle corresponding to $q_{n-1}^*$.
\item Uneven coverage: This strategy considers the direction of the beam variation within a time window. As shown in Fig.~\ref{fig: beamselection}, suppose the angle variation changes clockwise, we assign $|\mathcal{S}|-J_0$ for $J_0 \geq 1$ candidate beams at most to the clockwise direction including $q_{n-1}^*$ and reserve $J_0$ candidate beams in the counterclockwise direction to account for potential alignment errors due to noise. 
\item Interleaved coverage:  It is similar to the uneven coverage, except that the candidate beams are selected in an interleaved manner.
\end{itemize}

\subsection{Beam Prediction}\label{sec: prediction}

Our beam prediction module leverages ODE-LSTM proposed in~\cite{Ma2023}, which achieves high prediction accuracy using all available beams to collect the pilot signals. When partial candidate beams are used for prediction, i.e., in the beam tracking mode, we show that ODE-LSTM still accurately predicts the optimal beam. Due to the fewer beams used in the beam tracking mode, ODE-LSTM can be simplified, contributing to faster training and prediction speed. For completeness, we outline the critical components in ODE-LSTM and refer the interested readers to~\cite{Ma2023} for details.

\begin{figure}[!t]
\centering
{
\includegraphics[width=0.8\linewidth]{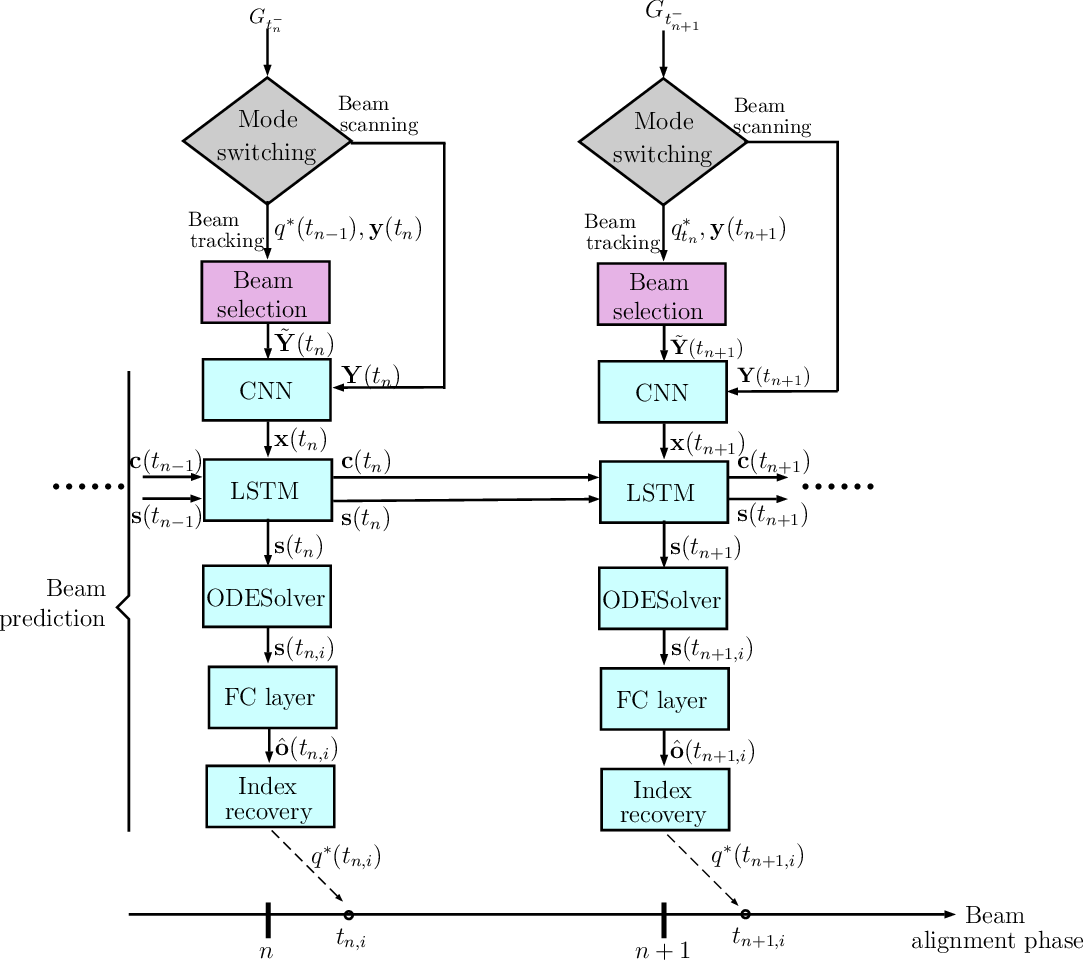}
}
\caption{The workflow of the proposed AOBA, where $t_{n,i}$ represent the $i$-th prediction instant between $t_n$ and $t_{n+1}$.}
\label{fig: flowchart}
\end{figure}

Fig.~\ref{fig: flowchart} depicts the operation procedure of the beam prediction module in the proposed AOBA. To illustrate, we assume that beam training is invoked in the $n$-th beam alignment stage. Given the received pilot signal vector denoted as $\tilde{\mathbf{Y}}(t_n)=[\mathbf{y}_{t_n}^{(q)}]_{q\in \mathcal{S}}$, a convolutional neural network (CNN) is used to extract the preliminary feature vector denoted as $\mathbf{x}(t_n)$, which is fed into a standard long short-term memory (LSTM) cell for further feature extraction. If beam scanning is performed in the $n$-th beam alignment stage, the received pilot signal is denoted as $\mathbf{Y}(t_n)=[\mathbf{y}_{t_n}^{(q)}]_{q=1,\cdots,Q}$. The core idea is that when the UE moves, its channel angle variations are correlated in time. Hence, the channel information of previous time steps can be used to predict the optimal beams of the future time steps based on LSTM. In particular, the forgetting and updating mechanisms in LSTM are useful to learn the channel variations, thereby improving beam prediction accuracy. A standard LSTM cell comprises a cell state $\mathbf{c}(t_n)$, a hidden state $\mathbf{h}(t_n)$, an input feature $\mathbf{x}(t_n)$, an input gate, an output gate, and a forget gate. The cell state remembers the previous data, and the hidden state enables the LSTM to learn non-linear dynamics. The forget gate determines which information should be omitted from memory at a specific time. The input gate consists of two components: the sigmoid function is responsible for determining which feature should be selected, and the other is the tanh function, which updates weights for the propagated values. Finally. the output gate is multiplied by the previous hidden state $\mathbf{s}(t_{n-1})$ and the current input $\mathbf{x}(t_n)$ to produce the output along with the new hidden state $\mathbf{h}(t_n)$ and the cell state $\mathbf{c}(t_n)$.
  
The standard LSTM can only provide a one-step prediction. To remedy this, an ODE solver is added to capture beam variation in continuous time~\cite{Ma2023}. Since the hidden state during the prediction duration is unknown, the ODE solver learns the derivative of the hidden state function with respect to time, i.e., $\mathrm{d}\mathbf{s}(t)/\mathrm{d}t$   Then by solving an initial value problem using an ODE solver, an approximation of the unknown hidden state at an arbitrary time instant can be obtained. While the ODE solver can find the hidden state continuously, it can also be implemented in discrete time. For example, if the prediction duration is divided into multiple time slots, we can find the hidden state of the $i$-th slot in the $n$-th prediction stage by using the ODE solver with the output denoted as $\mathbf{s}(t_{n,i})$. Finally, a fully connection (FC) layer is employed to transform the hidden state $\mathbf{s}(t_{n,i})$ into the candidate beam set. The output of the FC layer is normalized into the probabilities through the softmax activation layer. Let $\mathbf{o}(t_{n,i})=[o^{(1)}(t_{n,i}), o^{(2)}(t_{n,i}), \cdots, o^{(|\mathcal{S}|)}(t_{n,i})]$ the output of the activation layer where $o^{(q)}(t_{n,i})$ represents the probability of the $q$th beam in set $\mathcal{S}$ being the optimal one at time $t_{n,i}$.  Based on $o^{(q)}(t_{n,i})$, the beam with the maximal probability is selected as the optimal beam with the index given by
\begin{equation}\label{eq: optimal-beam-index}
\hat{q}^*(t_{n,i}) = \argmax_{q \in \mathcal{S} }~o^{(q)}(t_{n,i}).
\end{equation}
Notice that the beam index in \eqref{eq: optimal-beam-index} may differ from the actual beam index when $S<Q$. Fortunately, the true beam index can be quickly recovered based on the knowledge of the candidate beam set $\mathcal{S}$. 

\subsection{Mode Switching}\label{sec: mode-switching}

One limitation of beam prediction is that the prediction accuracy degrades with time. This problem can be alleviated by performing periodic beam training to get a timely update of the beam variation pattern. As mentioned, we consider two operation modes in the beam training stage: beam scanning using all available beams and beam tracking using partially selected candidate beams. By default, beam tracking is used for its low overhead. We propose two switching rules to switch from beam tracking to beam scanning mode. To ease our explanation, define $\zeta \in \{\texttt{True}, \texttt{False}\}$ a boolean variable. When $\zeta$ is \texttt{True}, mode switching occurs and $\zeta$ is \texttt{False} otherwise.
\begin{itemize}
\item Periodic switching: Switch from beam tracking to beam tracking in a fixed period denoted as $T_s$. The corresponding switching rule can be expressed as
\begin{equation}
\zeta = \begin{cases}
\texttt{True}, & \text{if}~t_n \bmod~T_s = 0,\\
\texttt{False}, & \text{otherwise}.
\end{cases}
\end{equation}
\item Adaptive switching: Switch from beam tracking to beam scanning according to the quality of the latest prediction result measured by the beamforming gain $G_{t^-_n}$, where $t^-_n$ represents the last prediction instant in the $n$-th prediction duration. If $G_{t^-_n}$ is less than a prescribed threshold denoted as $G_0$, it implies the prediction based on the candidate beam set $\mathcal{S}$ is no longer sustainable and complete beam scanning should be invoked. Otherwise, beam training can be used without change. Such an adaptive switching rule can be expressed as
\begin{equation}
\zeta = \begin{cases}
\texttt{True}, & \text{if}~G_{t^-_n}< G_0\\
\texttt{False}, & \text{otherwise}.
\end{cases}
\end{equation}
\end{itemize}
In comparing these two switching rules, periodic switching is an open-loop mechanism because it does not use any information about the previous prediction result. On the other hand, adaptive switching is closed-loop because the UE needs to report the thresholding result to the BS. Despite the extra overhead due to UE feedback, the adaptive switching is more robust to diverse beam variation scenarios and bypasses the difficulty of determining a proper switching period. We will compare their performance in Sec.~\ref{sec: results}.

\section{Simulation Results}\label{sec: results}

\begin{table}[H]
\caption{Layer parameters used for beam scanning and beam tracking (marked by $^\ast$)}
\label{table: training model}
\begin{center}
\footnotesize{
\begin{tabular}[b]{lll}
\hline
Layer & Output size & Activation \\
\hline
$3~(2^*) \times$ Conv2D	\quad\quad	&$32 \times 256 (^\ast 64) \times 10$ \quad\ &ReLU\\
LSTMcell				\quad\quad	&$32 \times 256 (^\ast 64) $		\quad	&Sigmoid,tanh
\\
ODESolver\quad\quad	& $32 \times 256 (^\ast 64)$		\quad	& tanh\\
FC\quad\quad	& $32 \times 64 (^\ast 11)$	\quad	&Softmax \\
\hline
\end{tabular}
}
\end{center}
\end{table}

In the simulation, we use DeepMIMO~\cite{Alkhateeb2019}, which has been widely used to generate mmWave channel datasets based on ray tracing. We consider the outdoor scenario (O1) and the first user grid (UG1). We collect samples of different UE locations following 10,240 trajectories to model user mobility. Details about the mobility modeWe implement the proposed  AOBA in PyTorch with the prediction module structure listed in Table~\ref{table: training model}. The simulation parameters follow those considered in~\cite{Ma2023}, except that the beam alignment period $T$ is 100 ms, which is also identical to the period of mode switching $T_s$. Two sets of evaluations are conducted in the following.
\begin{itemize}
\item The mode-switching module is deactivated so that we can identify the efficacy of the proposed beam selection strategy along with beam prediction based on ODE-LSTM. In this case, $|\mathcal{S}|=11$ candidate beams selected every $T$ seconds are used to predict the optimal beams for 99-time instants with uniform intervals in $T$ seconds.
\item Set B: The mode switching module is activated. Our goal is to examine the efficacy of the mode-switching module by performing the complete AOBA scheme.
\end{itemize}
We compare the performance of state-of-the-art beam alignment schemes with ours in terms of the normalized beamforming gain defined as
\begin{equation}\label{eq: normalized beamforming gain}
\bar{G}_t = \frac{ |\mathbf{H}_t^T \mathbf{f}_{\hat{q}^*(t)}|^2 }{ |\mathbf{H}_t^T \mathbf{f}_{q^*(t) }|^2 }
\end{equation} 
where $q^*(t)$ and $\hat{q}^*(t)$ denote the index of the true and the predicted optimal beam, respectively, at time $t$.

\begin{figure}[!t]
\centering
{
\includegraphics[width=0.7\linewidth]{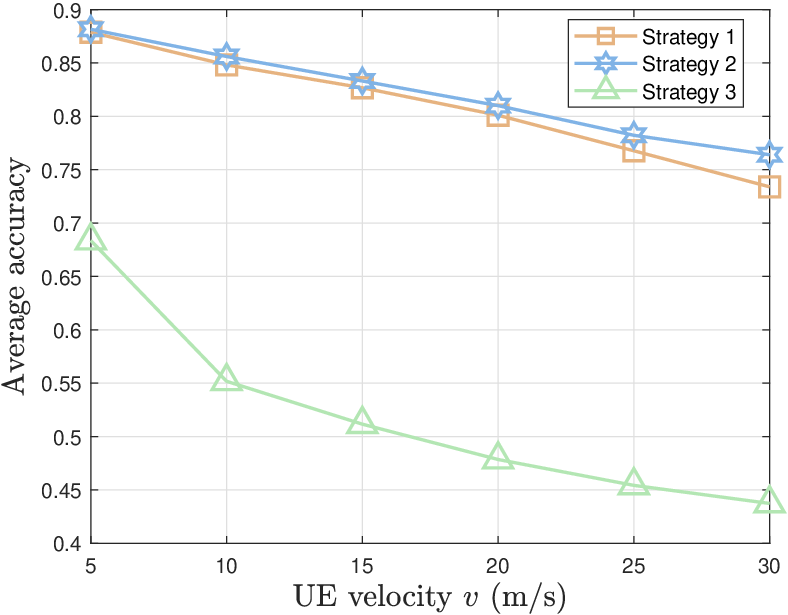}
}
\caption{Comparison of prediction accuracy of three candidate beam selection strategies versus UE velocity.}
\label{fig: comparison-strategy_accuracy-vs-velocity}
\end{figure}

\noindent \textit{Comparison of beam selection strategies (Set A)}: We first compare the prediction accuracy of three candidate beam selection strategies in Fig.~\ref{fig: comparison-strategy_accuracy-vs-velocity} with varied UE velocity and $J_0$=$2$. It can be seen that the uneven strategy and the even strategy significantly outperform the interleaved strategy. While the interleaved strategy also considers the moving direction as the uneven strategy, the interleaving magnifies the angular quantization errors of the analog beamformers. This results in a lower sampling rate of a time series that causes learning failure of LSTM. Since the uneven strategy outperforms the other two strategies, we will use it in the rest of the simulations.

\begin{figure}[!t]
\centering
\subfloat[Prediction performance within one $T$.]{\includegraphics[width=0.45\linewidth]{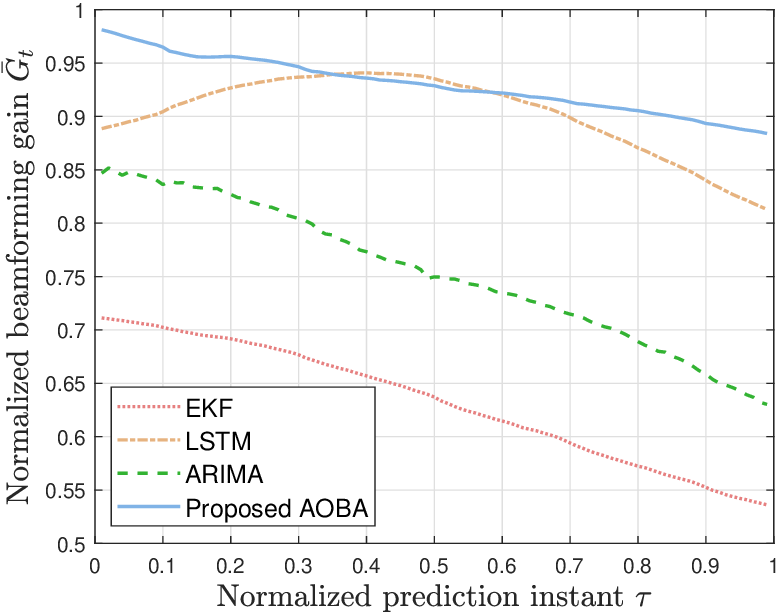}
\label{fig: G vs time}}
\subfloat[Impact of UE speed.]{\includegraphics[width=0.45\linewidth]{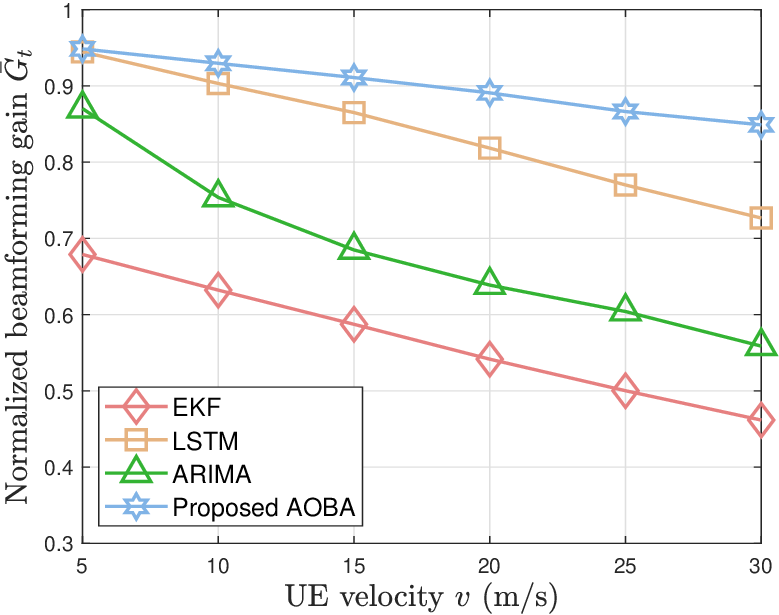}
\label{fig: G vs v}}
\caption{Comparison of different beam prediction methods for 1-second trajectories.}
\label{fig: results-1-seconds}
\end{figure}

\noindent \textit{Comparison of beam prediction methods (Set A)}: Next, we examine the prediction performance within one prediction period, considering three benchmarks: autoregression integrated moving average (ARIMA), LSTM, and extended kalman filter (EKF). They are commonly used to carry out time-series forecasting~\cite{SiamiNamini2018}, the same problem as beam prediction in our work. Except for ARIMA, which assumes linear dependence between observations, all the other methods can work for a nonlinear process. We use Python library PMDARIMA~\cite{PSF2023} to select the parameters of ARIMA, including the order of autoregression, degree of differencing, and the order of moving average. Based on the obtained parameters, ARIMA uses the predicted optimal beam indices of the previous four beam alignment stages and the optimal one of the current beam alignment stage to predict the optimal beams in $T$ seconds. Fig.~\ref{fig: G vs time} shows the normalized beamforming gain in the normalized prediction instant defined as $\tau=(t_{n_i}-t_n)/T$ for $i=1,2,\cdots,99$~\cite{Ma2023}. Here, we fix the UE velocity $v=10$ m/s, and each trajectory lasts 1 second. It is observed that the proposed AOBA consistently offers better accuracy than the other methods. The curve of LSTM shows a single peak because it can only deliver a one-step prediction, which tends to be the middle of $T$ to minimize the overall performance degradation, as explained in~\cite{Ma2023}. As to ARIMA, it outperforms EKF but performs much worse than LSTM and the proposed AOBA. From our experiment, ARIMA works well when trained to predict a single trajectory but becomes weak when trained to predict different trajectories. Finally, EKF is poorly performed because the error covariance matrix used in EKF to capture the prediction error can only be updated once per prediction period $T$ (using the received pilot signals), which leads to accumulated errors in time. 

In Fig.~\ref{fig: G vs v}, we show the impact of UE velocity to $\bar{G}_t$. At low speed, LSTM performs comparably to the proposed AOBA. LSTM incurs a significant degradation similar to ARIMA and EKF as the UE velocity increases. It is observed that $\bar{G}_t$ of the proposed AOBA decays slower than the competing approaches, demonstrating the robustness of AOBA to UE speed.
 
\begin{figure}[!t]
    \subfloat[Prediction performance within one $T$.]{%
      \includegraphics[width=0.5\linewidth]{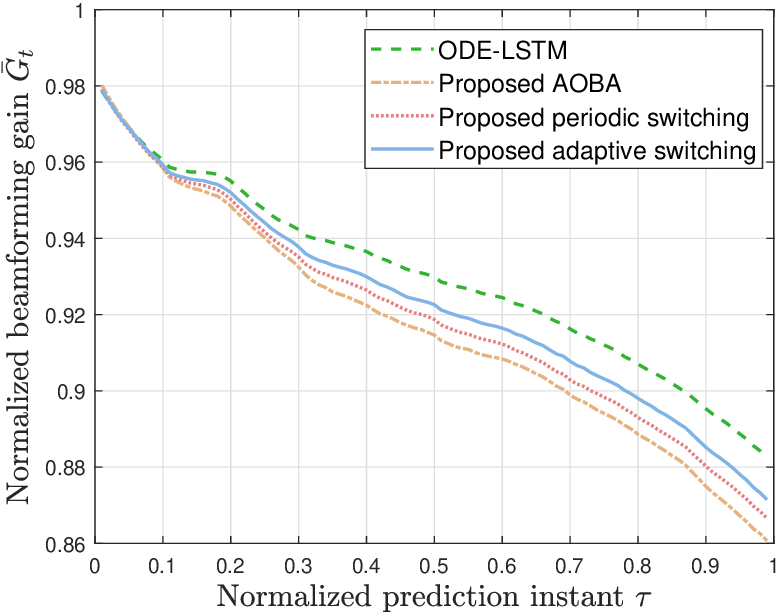}\label{fig: G vs time-switching}
    }
    \subfloat[Impact of UE speed.]{%
      \includegraphics[width=0.5\linewidth]{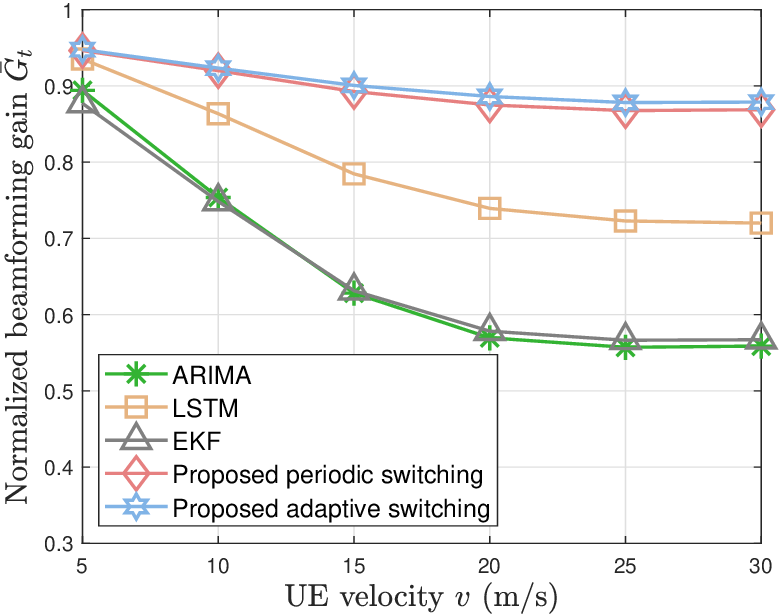}\label{fig: G vs v-switching}
    }
    \caption{Comparison of different beam prediction methods for 4-second trajectories.}
    \label{fig: results-4-seconds}
  \end{figure}

\noindent \textit{Comparison of mode switching rules (Set B)}: To evaluate the gain of mode switching, Fig.~\ref{fig: G vs time-switching} shows $\bar{G}_t$ in one prediction duration $T$ achieved by periodic switching and adaptive switching, where the switching threshold $G_0$ is computed by averaging the normalized beamforming gains of all the predicted optimal beams so far. Here, each trajectory is extended to 4 seconds. For comparison, we also include the results of AOBA without mode switching. Hence, ODE-LSTM serves as a performance upper bound because it makes predictions using the pilot signals over all available beams. The proposed AOBA with adaptive switching performs better than periodic switching and outperforms AOBA without mode switching.

Further, Fig.~\ref{fig: G vs v-switching} shows the impact of UE velocity to $\bar{G}_t$. Compared with the 1-second case in Fig.~\ref{fig: G vs v}, AOBA with mode switching maintains nearly the same $\bar{G}_t$ even when the trajectory length increases four times. On the contrary, other competing schemes become more sensitive to velocity increase, which can be seen from the faster decay of $\bar{G}_t$ as $v$ increases. The above results confirm the merit of mode switching in offering high prediction accuracy for a wide range of UE velocities.

Finally, we evaluate the normalized training overhead of various prediction schemes, where the training overhead is computed as the number of beams used in sending the pilot signals normalized to the training overhead using complete beam scanning every $T$ seconds. Except for AOBA, all the competing schemes, i.e., EKF, ARIMA, and LSTM, do not consider mode switching. Thus, they fall in the category labeled as ``Not enabled'' with identical training overhead in Table~\ref{tab: overhead} for $v=5\sim 30$ m/s. It can be seen that both the proposed mode-switching schemes incur the same amount of training overhead at the low-speed region. As the speed increases, the adaptive switching mode incurs a slightly higher overhead due to the increased frequency of invoking beam scanning. Compared with periodic mode switching, adaptive mode switching only has a $4\%$ increase of training overhead at $v=30$ m/s, and it is dramatically lower than the case of beam scanning. When mode switching is not enabled, 11 out of 64 (corresponding to $17\%$) beams are used per beam alignment period. Despite its low training overhead, the prediction accuracy degrades with UE velocity significantly, as shown in Fig.~\ref{fig: G vs v-switching}.

\begin{table}
\centering
\caption{Normalized training overhead}
\label{tab: overhead}
\footnotesize{
\begin{tabular}{ |c|c|c|c|c|}
\hline
\multicolumn{2}{|c|}{\multirow{2}{*}{Mode switching}} &  \multicolumn{2}{|c|}{Enabled} & \multirow{2}{*}{Not enabled} \\
\cline{3-4}
\multicolumn{1}{|c}{} & & Periodic & Adaptive & \\
\hline
\multirow{6}{*}{Speed} & 5 m/s &  \multirow{6}{*}{$52\%$} &  $52\%$ & \multirow{6}{*}{$17\%$}\\
\cline{4-4}
 & 10 m/s & &  $52\%$ &  \\
 \cline{4-4}
 & 15 m/s & &  $53\%$ & \\
 \cline{4-4}
 & 20 m/s  &   & $55\%$ & \\
 \cline{4-4}
 & 25 m/s & & $56\%$ & \\
 \cline{4-4}
 & 30 m/s & & $56\%$ & \\ 
\hline
\end{tabular}
}
\end{table}

\section{Conclusion}\label{sec: conclusion}
Compared with state-of-the-art beam prediction schemes, including ARIMA, EKF, and LSTM, our simulation results show that AOBA successfully reduces training overhead for accurate beam alignment in a single-user mmWave communication thanks to the proposed beam selection rule with the beam variation direction taking into account. When mode switching is enabled, AOBA with periodic mode switching performs comparably to the performance upper bound using ODE-LSTM with complete beam scanning while saving about $48\%$ of the training overhead. The adaptive mode switching incurs a slightly increased training overhead as the UE velocity increases due to more frequently invoked beam scanning. With the aid of mode switching, AOBA is more robust to UE velocity and consistently maintains the normalized beamforming gain close to $90\%$ when the trajectory length is increased from one second to four seconds in our experiment. The results reveal that the proposed AOBA can achieve high beam prediction accuracy using very low training overhead, which is vital to mobile mmWave applications. The extension to the multi-user beam alignment should deserve further research, where multiple UEs within the same BS coverage have different moving patterns that further challenge accurate beam prediction. The source codes for reproducing the results in this paper are available at~\cite{Lin2023}.

\section*{Acknowledgment}

This work was supported in part by the Ministry of Science and Technology, Taiwan, under Grant MOST 111-2221-E-007-076-MY3. We appreciate the authors in~\cite{Ma2023} for the offline discussions and sharing their simulation details.

\bibliographystyle{IEEEtran}
\bibliography{ref}

\end{document}